\documentstyle[twoside,fleqn,espcrc2]{article}

\expandafter\let\expandafter\originaleqnarray
  \csname eqnarray\endcsname % save the orig. def. of \eqnarray
\expandafter\def\csname eqnarray\endcsname
 {\settowidth{\arraycolsep}{$\mskip 0.5\thickmuskip$}\originaleqnarray}
\expandafter\let\expandafter\eqnarraystar
  \csname eqnarray*\endcsname % save the orig. def. of \eqnarray*
\expandafter\def\csname eqnarray*\endcsname
 {\settowidth{\arraycolsep}{$\mskip 0.5\thickmuskip$}\eqnarraystar}

\def\stupidskip{\!\!\!\!\!\!\!\!\!\!\!\!\!\!\!\!\!\!\!\!}

\let\a \alpha    \let\b \beta     \let\g \gamma    
\let\d \delta
\let\e \epsilon
      \let\h \eta      \let\q \theta

\let\l \lambda  \let\m \mu      \let\n \nu            
\let\x \xi

        \let\s \sigma              
                    
\let\f \phi
                   
\let\vf \varphi
\let\w \omega

\let\la \label    \let\nn \nonumber

      \let\del \partial      

\newcommand{\be}{\begin{equation}}
\newcommand{\ee}[1]{\label{#1}\end{equation}}
\newcommand{\bea}{\begin{eqnarray}}
\newcommand{\eea}{\end{eqnarray}}
\newcommand{\ra}{\rightarrow}

\newcommand{\NPB}[3]{Nucl.\ Phys. B#1 (#2) #3}
\newcommand{\CMP}[3]{Commun.\ Math.\ Phys.\ #1 (#2) #3}
\newcommand{\PRD}[3]{Phys.\ Rev.\ D#1 (#2) #3}
\newcommand{\PLB}[3]{Phys.\ Lett.\ B#1 (#2) #3}
\newcommand{\tmath}[1]{\mbox{$#1$}}

\newcommand{\refer}[1]{(\ref{#1})}
\newcommand{\wtd}[1]{\widetilde{#1}}
\newcommand{\qb}{\bar{\q}}
\newcommand{\lp}{{\l^\prime}}
\newcommand{\lpp}{{\l^{\prime\prime}}}
\newcommand{\lppp}{{\l^{\prime\prime\prime}}}
\newcommand{\whQ}{{\widehat{Q}}}
\newcommand{\ft}[2]{{\textstyle\frac{#1}{#2}}}

\title{\vspace{-3truecm}
{\small
\rightline{THU-97-44}
\rightline{NIKHEF 97-045}
\rightline{hep-th/9710215}}
\vspace{1truecm}
Open and Closed Supermembranes with Winding}
\author{Bernard de Wit\address{Institute for Theoretical Physics, Utrecht University\\
Princetonplein 5, 3508 TA Utrecht, The Netherlands\\
},
Kasper Peeters$^{\rm b}$ and Jan C.\ Plef\/ka\address{NIKHEF, 
P.O. Box 41882, 1009 DB Amsterdam, The Netherlands\\ 
}
}
\begin{document}
\begin{abstract}
Motivated by manifest Lorentz symmetry and a well-defined
large-$N$ limit prescription, we study the supersymmetric quantum
mechanics proposed as a model for the collective dynamics of 
D0-branes from the point of view of the 11-dimensional supermembrane. 
We argue that the continuity of the spectrum persists irrespective 
of the presence of winding around compact target-space directions 
and discuss the central charges in  
the superalgebra arising from winding membrane configurations.
Along the way we comment on the structure of open supermembranes.
\end{abstract}
\maketitle

M-theory is defined as the strong-coupling limit of type-IIA 
string theory and is supposed to capture all the relevant degrees 
of freedom of all known string theories, both at the perturbative and the 
nonperturbative level \cite{Townsend,witten3,horvw,Mtheory}. In 
this description the various string-string dualities play a  
central role. At large distances M-theory is described by 
11-dimensional supergravity. In \cite{bergs87} it was shown that 
elementary supermembranes can live in a superspace background 
that is a solution of the source-free supergravity field 
equations in 11 dimensions. In the light-cone gauge (in a 
flat target space) it was subsequently shown \cite{dWHN} that the 
supermembrane theory takes the  
form of a supersymmetric quantum-mechanical model, which 
coincides with
the zero-volume reduction of supersymmetric Yang-Mills 
theory based on 16 supercharges. For the supermembrane the 
underlying gauge group is  
the group of area-preserving diffeomorphisms of the 
two-dimensional membrane surface. This group can be described by 
the $N\to\infty$ limit of SU$(N)$ (the role of the membrane 
topology is subtle, as we will discuss in due course). For the 
finite groups the phase-space variables take the 
form of matrices, associated with the Lie algebra of some group 
(in this case U$(N)$ or SU$(N)$). For this reason, these models 
are commonly referred to as {\it matrix} models. It has been 
discussed that the  
possible massless ground states of the supermembrane coincide 
with the physical states of 11-dimensional supergravity\footnote{%
  For discussions on the existence of massless states, see 
  \cite{dWHN,DWN,FH,mboundstates}. According to \cite{mboundstates} 
  such states do indeed exist in eleven dimensions. }. 

More recently it was shown that the same quantum-mechanical 
matrix models based on U$(N)$ describe the short-distance dynamics of 
$N$ D0-branes \cite{boundst,Dbranes}. Subsequently there has been a 
large number of studies of these models for finite $N$ 
\cite{Dpart,BBPT} and some of them have been reported at this 
conference. These studies were further motivated by a conjecture 
according to which the degrees of freedom captured in M-theory, 
are in fact  
described by the U$(N)$ super-matrix models in the $N\to \infty$ 
limit \cite{BFSS}. A further conjecture, also discussed at this 
conference, is that the finite-$N$ matrix model coincides with 
M-theory compactified on a light-like circle \cite{susskind}.

So it turns out that M-theory, supermembranes and super-matrix 
models are intricately related. A direct relation between 
supermembranes and type-IIA theory was emphasized in particular in 
\cite{Townsend}, based on the relation between $d=10$ extremal
black holes in 10-dimensional supergravity and the Kaluza-Klein 
states of 11-dimensional supergravity. From the string 
point of view these states carry Ramond-Ramond charges, just as 
the D0-branes \cite{Polchinski}. Strings can 
arise from membranes by a so-called double-dimensional 
reduction \cite{DHIS}. Similarly supermembranes were employed
to provide evidence for the duality of M-theory on $\mbox{\bf R}^{10}\times
S_1/\mbox{\bf Z}_2$ and 10-dimensional $E_8\times E_8$ heterotic strings
\cite{horvw}.

Here we choose the supermembrane perspective, motivated 
by its manifest Lorentz invariance and well-defined large-$N$ 
limit (but not necessarily committing ourselves to the view that 
M-theory is a theory of fundamental membranes). Let us 
nevertheless first consider the supersymmetric  
matrix models, whose Hamiltonian equals   
\begin{equation}
H = \frac{1}{g}{\rm Tr}\Big[ \ft{1}{2}{\bf P}^2 + \ft{1}{4}[X^a,X^b]^2
+ g\theta^{\rm T} \gamma_a [ X^a, \theta ]\Big] \, .
\end{equation}
Here, $\bf X$, $\bf P$ and $\theta$ take values in the Lie 
algebra of the gauge group. From the supermembrane point of view 
they are vectors and spinors of the `transverse' SO(9) rotation
group. This Hamiltonian can alternatively be interpreted as the
zero-volume limit of supersymmetric Yang-Mills theory with some 
arbitrary gauge group. For the supermembrane one 
must choose the (infinite-dimensional) group of area-preserving
diffeomorphisms (which also plays an important role in selfdual
gravity and $N=2$ strings) and put $g$ equal to the light-cone momentum 
$P^+_0$. 
The matrix model has 16 supercharges, but additional charges can 
be obtained by  splitting off an abelian factor of the U$(N)$ 
gauge group, 
\begin{eqnarray}
Q^+ &=& {\rm Tr} \Big[(2P^a \gamma_a + [X^a,
X^b]\gamma_{ab})\theta\Big]\,,  \nonumber\\ 
Q^- &=& {g}\; {\rm Tr} \left[ \theta \right] \, .
\end{eqnarray}
For the supermembrane the second charge is associated with the 
center-of-mass superalgebra (we return to the membrane 
supersymmetry algebra shortly). 

The form of the area-preserving diffeomorphisms that remain as an
invariance of the model, depends in general on the topology of the
membrane surface. For spherical and toroidal topologies, it has been
shown that the algebra can be approximated by SU$(N)$ in the 
large-$N$
limit. This limit is subtle. However, once one assumes an 
infinitesimal gauge group, one can describe a large variety of 
different theories. For instance, the gauge group $[{\rm 
U}(N)]^M$, which is a subgroup of U$(N\cdot M)$, leads in the 
limit $M\to \infty$ to the 
possibility of describing the collective dynamics of D0-branes on 
a circle by supersymmetric Yang-Mills
theories in $1+1$ dimensions \cite{Tdual}. Hence, it is possible 
to extract extra dimensions from a suitable infinite-dimensional 
gauge group. Obviously  
this can be generalized to a hypertorus. Therefore models based 
on (subgroups of) U$(\infty)$ can describe an enormous variety of 
models.  

Using the SU$(N)$ regularisation it was shown in \cite{dWLN} that the
spectrum of the supermembrane is continuous, with no mass gap. This
result is expected when the supermembrane Hamiltonian is viewed as the
Hamiltonian for the collective dynamics of arbitrarily large numbers
of D-particles. From the supermembrane point of view, these
instabilities arise because arbitrarily long, string-like (zero-area)
configurations can form. There are thus no asymptotic membrane states,
but rather multimembrane configurations connected by these infinitely
thin strings. The massless ground states, which correspond to the 
states of 11-dimensional supergravity, thus appear in a continuum 
of multimembrane states. 

The connection with M-theory and finite-N matrix models is made by
compactification of the 11-dimensional action on a circle or 
higher-dimensional tori. However, this compactification has only recently
been studied \cite{us} from the point of view of the supermembrane.
In this talk we 
discuss the supermembrane with winding, paying attention to the
extension of the supersymmetric gauge theory, the Lorentz invariance
and the effect of winding on the mass spectrum. Along the way we
shall simultaneously develop the theory of open supermembranes,
which has recently received some attention \cite{BB,BM,EMM2}.
                                                          
The actions of fundamental supermembranes are of the Green-Schwarz type
\cite{bergs87}. In flat target space, they read 
\bea\label{action}
{\cal L}&=&\sqrt{-g(X,\theta)} \nonumber\\
&& -\epsilon^{ijk}\, [ \ft{1}{2}\partial_iX^\mu
\, (\partial_jX^\nu+\qb \g^\nu\partial_j\q)\nn\\
&&\hspace{12mm} 
+\ft{1}{6}\,\qb\g^\mu\partial_i\q\, \qb\g^\nu\partial_j\q]\, \qb\g_{\mu\nu}
\partial_k\q\, , 
\eea
where $X^\mu$ 
denote the  11-dimensional
target-space embedding coordinates lying in 
$T^d\times{\mbox{\bf R}}^{1,10-d}$ 
and thus permitting us to have winding on the $d$-dimensional 
torus $T^d$. Moreover we have the fermionic variables $\q$, 
which are 32-component Majorana spinors and $g=\mbox{det}\, g_{ij}$ with
the induced metric
\be
g_{ij}=(\partial_iX^\mu+ \qb\gamma^\mu\partial_i\q)\,
   (\partial_jX^\nu+ \qb\gamma^\nu\partial_j\q)\, \eta_{\mu\nu}.
\ee{indmet}
Next to supersymmetry the action \refer{action} exhibits an additional
local fermionic symmetry called $\kappa$-symmetry. In the case of
the open supermembrane, $\kappa$-symmetry imposes boundary conditions
on the fields. They must ensure that the following integral over the 
boundary of the membrane world volume vanishes,
\bea
\int_{\del M} \Big[&& \ft12{\rm d} X^\m \wedge ( {\rm d}X^\n + 
\bar\theta\g^\nu {\rm d}\theta)\, \bar \theta \g_{\m\n} \d_\kappa 
\theta \nn\\
&&+\ft12( {\rm d}X^\m - \ft13 \bar\theta\g^\mu {\rm 
d}\theta ) \wedge 
\bar\theta\g_{\m\nu} {\rm d}\theta\; \bar\theta\g^\n \d_{\kappa}\theta
\nn \\[1mm]
&& +\ft16\bar\theta\g^\mu {\rm d}\theta\wedge \bar\theta\g^\nu {\rm 
d}\theta\, \bar\theta\g_{\m\nu} \d_{\kappa}\theta\Big ]\,.
\eea
This can be achieved by having a ``membrane D-$p$-brane'' at 
the boundary with $p=1,5$, or 9, which is defined in terms of 
$(p+1)$ Neumann and $(10-p)$ Dirichlet boundary 
conditions for the $X^\mu$, together with corresponding boundary conditions on 
the fermionic coordinates\footnote{%
  Here our conclusions concur with those of \cite{EMM2} but not with 
  those of \cite{BM}.}. % 
More explicitly, we define projection operators
\be 
{\cal P}_\pm=\ft12\Big({\bf 1} \pm \g^{p+1}\, \g^{p+2}\cdots 
\g^{10}\Big)\,, 
\ee{projectors}
and impose the Dirichlet boundary conditions
\bea
\del_\parallel \, X^M\big|&=& 0\,, \qquad M=p+1,\ldots,10\,, \nn\\
{\cal P}_- \q\big|&=&0\, , \la{boundcond}
\eea 
where $\del_\perp$ and $\del_\parallel$ define the world-volume 
derivatives perpendicular or tangential to the surface swept out 
by the membrane boundary in the target space. Note that the fermionic 
boundary condition implies that ${\cal P}_- \del_\parallel\q=0$. 
Furthermore, it implies that spacetime 
supersymmetry is reduced to only 16 supercharges associated with 
spinor parameters ${\cal P}_+\epsilon$, which is {\it chiral} with 
respect to the ($p+1$)-dimensional world volume of the 
D-$p$-brane at the boundary. With respect to 
this reduced supersymmetry, the superspace coordinates decompose 
into two parts, one corresponding to $(X^M, {\cal P}_-\theta)$ and the 
other corresponding to $(X^m, {\cal P}_+\theta)$ where
$m=0,1,\ldots,p$. While for the 
five-brane these superspaces exhibit a somewhat balanced decomposition in 
terms of an equal number of bosonic and fermionic coordinates,
the situation for $p=1,9$ shows heterotic features in that 
one space has an excess of fermionic and the other an excess of 
bosonic coordinates. Moreover, we note that supersymmetry may be further
broken by e.g.\ choosing different Dirichlet conditions 
on non-connected segments of the supermembrane boundary.

The Dirichlet boundary conditions can be supplemented by the 
following Neumann boundary conditions,
\bea
\del_\perp \, X^m\big|&=& 0 \qquad m=0,1,\ldots,p \,,\nn\\
{\cal P}_+ \del_\perp \q \big|&=&0 \,. \la{Nboundcond}
\eea 
These do not lead to a further breakdown of the rigid spacetime 
symmetries.

We now continue and follow the light-cone quantization described 
in \cite{dWHN}. The supermembrane Hamiltonian takes the form 
\begin{eqnarray}
\label{memham}
\displaystyle {\cal H}&=& \displaystyle\frac{1}{P_0^+}\, 
\int {\rm d}^2\s \, \sqrt{w}\,
\bigg[ \, \frac{P^a\, P_a }{2\,w} + \ft{1}{4} \{\, 
X^a,X^b\,\}^2\nonumber\\
\displaystyle &&\hspace{18mm}\quad\quad \displaystyle-P^+_0\, \qb\,\g_- 
\g_a\, \{\, X^a , \q\,\}\, \bigg]\, .
\end{eqnarray}
Here the integral runs over the spatial components of the
worldvolume denoted by $\s^1$ and $\s^2$, while 
$P^a(\s)$ ($a=2,\ldots,10$) are the momentum conjugates to the transverse
$X^a$. In this gauge 
the light-cone coordinate $X^+=(X^1+X^0)/\sqrt2$ is linearly related to the 
world-volume time. The momentum $P^+$ is time independent and 
proportional to the center-of-mass value $P^+_0$ times some 
density ${\sqrt{w(\s)}}$ of the spacesheet, whose spacesheet 
integral is normalized to unity. The center-of-mass momentum
$P_0^-$ is equal to minus the Hamiltonian \refer{memham}
subject to the gauge 
condition \tmath{\g_+\, \q=0}. And finally we made use of the
Poisson bracket \tmath{\{ A,B\} } defined by
\be
\{ A(\s ),B(\s )\} = \frac{1}{\sqrt{w(\s)}}\, \e^{rs}\, \del_r A(\s )\,
\del_s B(\s ).
\ee{poisbrak}
Note that the coordinate $X^-=(X^1-X^0)/\sqrt2$ itself does
not appear in the Hamiltonian \refer{memham}. It is defined via 
\be
P^+_0\, \del_rX^-= - \frac{{\bf P} \cdot \del_r{\bf X}}{\sqrt{w}} - 
P^+_0\, \qb\g_-\del_r\q\,,
\ee{delxminus}
and implies a number of constraints that will be important in 
the following. First of all, the right-hand side must be closed.
If there is no winding in $X^-$, it must moreover be exact.

The equivalence of the large-$N$ limit of SU$(N)$ quantum mechanics
with the closed supermembrane model is based on the residual invariance
of the supermembrane action in the light-cone gauge. It is given
by the area-preserving diffeomorphisms of the membrane surface.
These are defined by transformations of the worldsheet coordinates
\begin{equation}
\label{APD}
\s^r \ra \s^r + \x^r(\s) \,, 
\end{equation}
with
\begin{equation}
\del_r(\sqrt{w(\s)}\, \x^r(\s)\, )=0.
\end{equation}
We wish to rewrite this condition in terms of dual spacesheet 
vectors by  
\be                                  
\sqrt{w(\s)}\,\x^r(\s)= \e^{rs}\, F_s(\s)\, .
\ee{1form}
In the language of differential forms the
condition \refer{APD} may then be simply recast as \tmath{{\rm 
d}F=0}. The trivial solutions are the exact forms \tmath{F={\rm d}\x}, 
or in components 
\be
F_s=\del_s\x(\s),
\ee{exact}
for any globally defined function $\x(\s)$. The nontrivial solutions are
the closed forms which are not exact. On a Riemann surface of
genus $g$ there are precisely $2g$ linearly independent non-exact 
closed forms, whose integrals along the homology cycles are 
normalized to unity\footnote{%
  In the mathematical literature the globally defined exact forms 
  are called ``hamiltonian vector fields'', whereas the closed 
  but not exact forms which are not globally defined go under the 
  name ``locally hamiltonian vector fields''.}. %
In components we write
\be
F_s=\f_{(\l)\, s}\;, \qquad \l=1,\ldots,2g\,.
\ee{harm}

The commutator of two infinitesimal area-preserving 
diffeomorphisms is determined by the product rule
\begin{equation}
\xi_r^{(3)} = \partial_r \left( \frac{\epsilon^{st}}{\sqrt{w}} 
\xi_s^{(2)}\xi_t^{(1)}\right) \,,
\end{equation}
where both $\xi_r^{(1,2)}$ are closed vectors. Because 
$\xi_r^{(3)}$ is exact, the exact vectors 
generate an invariant subgroup of the area-preserving 
diffeomorphisms, 
which can be approximated by SU$(N)$ in the large-$N$ limit in the
case of closed membranes. For open membranes the boundary conditions 
on the fields \refer{boundcond} lead to an SO($N$) group structure, 
as we shall see in the sequel.

The presence of the closed but non-exact forms is crucial for 
the winding of the embedding coordinates. More precisely, while 
the momenta ${\bf P}(\s)$ and the fermionic coordinates 
$\theta(\s)$ remain single valued on the spacesheet, the 
embedding coordinates, written as one-forms with components 
$\del_r {\bf X}(\s)$ and  $\del_r X^-(\s)$, are decomposed into 
closed forms. Their non-exact contributions are multiplied by an 
integer times the length of the compact direction.
The constraint alluded to above 
amounts to the condition that the right-hand side of 
\refer{delxminus} is closed. 

Under the full group of area-preserving diffeomorphisms the fields $X^a$,
$X^-$ and $\q$ transform according to
\begin{eqnarray}
\label{APDtrafoXtheta}
&\d X^a= \displaystyle{\e^{rs}\over \sqrt{w}}\, \x_r\, \del_s X^a\,,
\quad 
\d X^-= \displaystyle {\e^{rs}\over \sqrt{w}}\, \x_r\, \del_s 
X^-\,,\nonumber\\ 
&\quad\quad\d \q^a= \displaystyle {\e^{rs}\over \sqrt{w}}\, 
\x_r\, \del_s \q\,, 
\end{eqnarray}
where the time-dependent reparametrization $\x_r$ consists of
closed exact and non-exact parts. Accordingly there is a gauge
field $\w_r$, which is therefore closed as well, transforming as
\be
\d\w_r=\del_0\x_r + \del_r \bigg( {\e^{st}\over\sqrt{w}}\,\x_s\,\w_t\bigg), 
\ee{APDtrafoomega}
and corresponding covariant derivatives
\begin{eqnarray}
\label{covderiv}
D_0 X^a&=& \displaystyle\del_0X^a - {\e^{rs}\over \sqrt{w}}\, 
\w_r\, \del_s X^a\,, \nonumber\\ 
D_0 \q &=&\displaystyle \del_0\q - {\e^{rs}\over \sqrt{w}}\, 
\w_r\, \del_s\q\,, 
\end{eqnarray}
and similarly for \tmath{D_0 X^-}. 

The action corresponding to the following Lagrangian density is
then gauge invariant under the 
transformations \refer{APDtrafoXtheta} and \refer{APDtrafoomega},
\bea
\label{gtlagrangian}
{\cal L}&=&P^+_0\,\sqrt{w}\, \Big[\,  
\ft{1}{2}\,(D_0{\bf X})^2 + \qb\,\g_-\,
D_0\q \nonumber\\
&& \hspace{13mm} - \ft{1}{4}\,(P^+_0)^{-2}\,  \{ X^a,X^b\}^2 \\
&& \hspace{13mm}  + (P^+_0)^{-1}\, \qb\,\g_-\,\g_a\,\{X^a,\q\} +
  D_0 X^-\Big]\, ,\nn 
\eea
where we draw attention to the last term proportional to
$X^-$, which can be dropped in the absence of winding and did not
appear in \cite{dWHN}. Moreover, we note that for the open
supermembranes, \refer{gtlagrangian} is invariant under the 
transformations \refer{APDtrafoXtheta} and \refer{APDtrafoomega} 
only if $\xi_\parallel=0$ holds on the boundary.
This condition defines a subgroup of the 
group of area-preserving transformations, which is consistent 
with the Dirichlet conditions \refer{boundcond}. Observe that 
$\del_\parallel$ and $\del_\perp$ will now refer to the {\it 
spacesheet} derivatives tangential and perpendicular to the 
membrane boundary\footnote{%
  Consistency of the Neumann boundary conditions 
  \refer{Nboundcond} with the area-preserving diffeomorphisms 
  \refer{APDtrafoXtheta} further imposes 
  $\partial_\perp\xi^\parallel=0$ 
  on the boundary, where indices are raised according to 
  \refer{1form}.}. % 

The action corresponding to
\refer{gtlagrangian} is also invariant under the 
supersymmetry transformations  
\bea
\d X^a &=& -2\, \bar{\e}\, \g^a\, \q\,, \nn\\
\d \q  &=& \ft{1}{2} \g_+\, (D_0 X^a\, \g_a + \g_- )\, \e
\nonumber\\
&&\quad\quad +\ft{1}{4}(P^+_0)^{-1} \,
\{ X^a,X^b \}\, \g_+\, \g_{ab}\, \e ,\nn\\
\d \w_r &=& -2\,(P^+_0)^{-1}\, \bar{\e}\,\del_r\q\,.
\la{susytrafos}
\eea
The supersymmetry variation of $X^-$ is not relevant and may be
set to zero. For the open case one finds that the boundary conditions
$\omega_\parallel=0$
and \mbox{$\epsilon={\cal P}_+\,\epsilon$} must be fulfilled
in order for \refer{susytrafos} to be a symmetry of the action.
In that case the theory takes the form of a gauge theory coupled 
to matter. The pure gauge theory is associated with the Dirichlet 
and the matter with the Neumann (bosonic and 
fermionic) coordinates.  

In the case of a `membrane D-$9$-brane' one now sees that the
degrees of freedom  on the `end-of-the world' $9$-brane precisely
match those of 10-dimensional heterotic strings. {\it On} the boundary
we are left with eight propagating bosons $X^m$ (with $m=2,
\ldots,9$), as $X^{10}$ is constant on the boundary 
due to \refer{boundcond},
paired with the 8-dimensional chiral spinors $\theta$ (subject 
to $\g_+ \theta= {\cal P}_-\theta=0$),
i.e., the scenario of Ho\u{r}ava-Witten \cite{horvw}.

The full equivalence with the membrane Hamiltonian is now established by
choosing the \tmath{\w_r=0} gauge and passing to the Hamiltonian 
formalism. The field equations for $\w_r$ then lead to
the membrane constraint \refer{delxminus} (up to exact contributions), 
partially defining \tmath{X^-}.
Moreover the Hamiltonian corresponding to the gauge theory Lagrangian of 
\refer{gtlagrangian} is nothing
but the light-cone supermembrane Hamiltonian \refer{memham}.
Observe that in the above gauge theoretical construction the space-sheet
metric $w_{rs}$ enters only through its density $\sqrt{w}$ and hence
vanishing or singular metric components do not pose problems.

We are now in a position to study the full 11-dimensional supersymmetry
algebra of the winding supermembrane. For this we decompose the
supersymmetry charge $Q$ associated with the transformations 
\refer{susytrafos}, into two 16-component spinors,
\be
Q= Q^+ + Q^- , \quad \mbox{where}\quad
 Q^\pm = \ft{1}{2}\, \g_\pm\,\g_\mp\, Q, 
\ee{Qdecomposition}
to obtain
\bea
Q^+&=&\int {\rm d}^2 \s \, \Big(\, 2\, P^a\, \g_a + \sqrt{w}\, \{\,
X^a, X^b\, \} \, \g_{ab}\, \Big) \, \q \,, \nn \\
Q^-&=& 2\, P^+_0\, \int {\rm d}^2\s\, \sqrt{w}\, \g_-\, \q .
\la{Q-cont}
\eea
The canonical Dirac brackets are derived by the standard
methods and read
\bea
(\, X^a(\s), P^b(\s^\prime)\, )_{\mbox{\tiny DB}} &=& \d^{ab}\, 
\d^2(\s-\s^\prime)\,, \nn\\
(\, \q_\a(\s), \qb_\b(\s^\prime)\, )_{\mbox{\tiny DB}} &=& \nonumber\\
&&\stupidskip\stupidskip
\ft{1}{4}\,(P^+_0)^{-1} \,w^{-1/2} \, (\g_+)_{\a\b}\,\d^2(\s-\s^\prime)\,.
\eea
In the presence of winding the results given in \cite{dWHN} yield the
supersymmetry algebra
%
% The most ugly LaTeX kludge you'll ever see in your life:
%
\begin{eqnarray}
\la{contsusy}
(\, Q^+_\a, \bar{Q}^+_\b\, )_{\mbox{\tiny DB}} &=& 2\, (\g_+)_{\a\b}\, {\cal H}\nonumber\\
 &&\stupidskip
 - 2\,  (\g_a\, \g_+)_{\a\b}\, \int{\rm d}^2\s\, \sqrt{w}\, \{\, X^a, X^-\,\}\, , \nn \\
(\, Q^+_\a, \bar{Q}^-_\b\, )_{\mbox{\tiny DB}} &=& -(\g_a\,\g_+\,\g_- )_{\a\b}\, P^a_0 \nonumber\\
 &&\stupidskip
  - \ft{1}{2}\,(\g_{ab}\, \g_+\g_- )_{\a\b}\, \int {\rm d}^2\s\,\sqrt{w}\, \{\, X^a,X^b\,\}\,,\nn\\
(\, Q^-_\a, \bar{Q}^-_\b\, )_{\mbox{\tiny DB}} &=& -2\, (\g_- )_{\a\b}\, P^+_0\, , 
\end{eqnarray}
where use has been made of the defining equation
\refer{delxminus} for $X^-$. 

The new feature of this supersymmetry algebra is the emergence of the 
central charges in the first two anticommutators, which are
generated through the winding contributions.
They represent topological quantities obtained by integrating
the winding densities 
\begin{equation}
z^{a}(\s)=\e^{rs}\,\del_r X^a\,\del_s X^-
\end{equation}
and
\begin{equation}
z^{ab}(\s) =\e^{rs}\,\del_r X^a\,\del_s X^b
\end{equation}
over the space-sheet. It is gratifying to observe the manifest
Lorentz invariance of \refer{contsusy}. Here  we should point out
that, in adopting the light-cone gauge, we assumed that there was
no winding for 
the coordinate $X^+$. In \cite{BSS} the corresponding algebra for
the matrix regularization was studied. 
The result obtained in \cite{BSS} coincides with ours in the
large-$N$ limit, in which an additional longitudinal five-brane
charge vanishes, provided that one identifies the longitudinal
two-brane charge with the central charge in the
first line of \refer{contsusy}. This requires the definition of
$X^-$ in the matrix regularization, a topic that was dealt with in
\cite{dWMN}. We observe that the discrepancy noted in \cite{BSS}
between the matrix calculation and 
certain surface terms derived in \cite{dWHN}, seems to have no
consequences for the supersymmetry algebra. A possible reason for
this could be that certain Schwinger terms have not been treated
correctly in the matrix computation, as was claimed in a recent
paper \cite{EMM}. 

The form of the algebra is another indication of the consistency
of the supermembrane-supergravity system.

In order to define a matrix approximation one introduces a complete 
orthonormal basis of functions $Y_A(\s)$ for the globally defined
$\x(\s)$ of \refer{exact}. One may then write down the following
mode expansions for the phase space variables of the
supermembrane, 
\bea
\del_r{\bf X}(\s) &=& {\bf X}^\l\, \f_{(\l)\, r} + \sum_A {\bf X}^A\, 
\del_r Y_A(\s),\nn \\
{\bf P}(\s) &=& \sum_A \sqrt{w}\, {\bf P}^A\, Y_a(\s), \nn\\
\q(\s) &=& \sum_A \q^A\, Y_A(\s) , \la{modeexp}
\eea
introducing winding modes for the transverse $X^a$. A similar expansion 
exists for $X^-$.  One then naturally
introduces the structure constants of the group of area-preserving 
diffeomorphism by \cite{dWMN}
\bea
f_{ABC} &=& \int {\rm d}^2\s\, \e^{rs}\,\del_r Y_A\, \del_s Y_B\,
Y_C\,, \nn \\ 
f_{\l BC} &=& \int {\rm d}^2\s\, \e^{rs}\,\f_{(\l)\, r}\, \del_s
Y_B\, Y_C\,, \nn \\ 
f_{\l \lp C} &=& \int {\rm d}^2\s\, \e^{rs}\,\f_{(\l)\, r}\,
\f_{(\lp)\, s}\, Y_C \, . 
\eea
Note that with $Y_0=1$, we have \tmath{f_{AB0}=f_{\l B0}=0}.
The raising and lowering of the $A$ indices is performed with the 
invariant metric
\begin{equation}
\h_{AB}= \int {\rm d}^2\s\, \sqrt{w}\, Y_A(\s)\, Y_B(\s)
\end{equation}
and there is no need to introduce a metric for the $\l$ indices.

By plugging the mode expansions \refer{modeexp} into the Hamiltonian 
\refer{memham} one obtains the decomposition
\bea
{\cal H} &=& \ft{1}{2}\, {\bf P}_0\cdot{\bf P}_{0}\nonumber\\
&&+ \ft{1}{4}\, {f_{\l \lp}}^0\, f_{\lpp \lppp 0}\, X^{a\, \l}\, X^{b\, \lp}\, 
X^\lpp_a\, X^\lppp_b \nn \\
&&+ \ft{1}{2}\, {\bf P}^{A}\cdot {\bf P}_{A} - f_{ABC}\, \qb^C\, \g_-\,\g_a\,
\q^B\, X^{a\, A} \nonumber\\
&& - f_{\l BC}\, \qb^C\, \g_-\,\g_a\,\q^B\, X^{a\, \l} \nn\\
&& + \ft{1}{4}\, {f_{AB}}^E\, f_{CDE}\, X^{a\, A}\, X^{b\, B}\, X^C_a\, X^D_b \nn\\
&&+ {f_{\l B}}^E\, f_{CDE}\, X^{a\, \l}\, X^{b\, B}\, X^C_a\, X^D_b \nn\\
&& +\ft{1}{2}\,  {f_{\l B}}^E\, f_{\lp DE}\, X^{a\, \l}\, X^{b\, B}\, 
X^\lp_a\, X^D_b \nn\\
&&+\ft{1}{2}\,  {f_{\l B}}^E\, f_{C \lp E}\, X^{a\, \l}\, 
X^{b\, B}\, X^C_a\, X^\lp_b \nn\\
&& + \ft{1}{2}\,  {f_{\l \lp}}^E\, f_{C DE}\, X^{a\, \l}\, X^{b\, \lp}\, 
X^C_a\, X^D_b \nn\\
&&+ {f_{\l \lp}}^E\, f_{\lpp DE}\, X^{a\, \l}\, X^{b\, \lp}\, 
X^\lpp_a\, X^D_b \nn\\
&& + \ft{1}{4}\, {f_{\l \lp}}^E\, f_{\lpp \lppp E}\, X^{a\, \l}\, X^{b\, \lp}
\, X^\lpp_a\, X^\lppp_b,
\la{memhammode}
\eea
where here and henceforth we spell out the zero-mode dependence
explicitly, i.e.\ the range of values for $A$ no longer includes
$A=0$. Note that for the toroidal supermembrane 
\tmath{f_{\l\lp A}=0} and thus the last three terms in \refer{memhammode}
vanish. The second term in the first line
represents the winding number squared. In the matrix formulation,
the winding number takes the form of a trace over a commutator.
We have scaled the Hamiltonian by a 
factor of $P^+_0$ and  the fermionic variables by a factor $(P^+_0)^{-1/2}$.
Supercharges will be rescaled as well, such as to eliminate
explicit factors of $P^+_0$. 

The constraint equation \refer{delxminus} is translated into mode
language by contracting it with \tmath{\e^{rs}\, \f_{(\l)\, s}} and
\tmath{\e^{rs}\, \del_s Y_C} respectively and integrating the result
over the spacesheet to obtain the two constraints
\bea
\vf_\l &=& f_{\l\lp 0}\,(\, {\bf X}^\lp\cdot{\bf P}_0 +X^{-\, \lp}\, P^+_0\, )
\nonumber\\
&&\quad+ f_{\l\lp C}\, {\bf X}^\lp\cdot{\bf P}^C \nn \\
&&\quad+ f_{\l BC}\,(\,  {\bf X}^B\cdot{\bf P}^C +\qb^C\,\g_-\, \q^B\, )
=0 ,\nn \\
\vf_A &=& f_{ABC}\, (\, {\bf X}^B\cdot{\bf P}^C + \qb^C\,\g_-\,\q^B\, )
\nonumber\\
&&\quad + f_{A\l C} \, {\bf X}^\l\cdot{\bf P}^C =0  ,
\eea
taking also possible winding in the $X^-$ direction into account.
Note that even for the non-winding case \tmath{X^{a\,\l}=0}
there remain the extra $\vf_\l$ constraints. These have so far
not played any role in the matrix formulation.

The zero-mode contributions completely decouple in the
Hamiltonian and the supercharges. We thus perform a split in
$Q^+$ treating zero modes and fluctuations separately to obtain
the mode expansions,  
\be
Q^- = 2\,\g_-\, \q^0 \,, \qquad Q^+ = Q^+_{(0)} + \whQ^+\,, 
\ee{split}
where
\bea
Q^+_{(0)} &=&  \Bigl (\,2\, P^a_0\,\g_a + f_{\l\lp 0}\,
X^{a\,\l}\, X^{b\,\lp}\, \g_{ab}\, \Bigr )\, \q_0 \,, \nn \\
\widehat{Q}^+ &=& \Bigl (\, 2\, P^a_C\,\g_a + 
f_{ABC}\, X^{a\, A}\, X^{b\, B}\,\g_{ab} \nn\\
&&\quad + 2\, f_{\l BC}\, X^{a\, \l}\, X^{b\, B}\,\g_{ab}\nn\\
&&\quad + f_{\l\lp C}\, X^{a\, \l}\, X^{b\, \lp}\,\g_{ab}\,  
\Bigr )\, \q^C \,.
\eea
Upon introducing the supermembrane mass operator by
\be 
{\cal M}^2=2\, {\cal H} - {\bf P}_0\cdot{\bf P}_{0}- \ft{1}{2}\,(
f_{\l\lp 0}\, X^{a\, \l}\, X^{b\, \lp})^2 ,
\ee{mass2}
the supersymmetry algebra \refer{contsusy} then takes the form
\bea
&&\{\whQ{}^+_\a, \bar{\widehat{Q}}{}^+_\b\, \} = (\g_+)_{\a\b} \, 
{\cal M}^2 \label{s1}\\
&&\hspace{4mm} -2\, (\g_a\,\g_+ )_{\a\b}\, f_{\l\lp 0}\, X^{a\,\l}\, ( X^{-\, 
\lp}\, P^+_0 +  {\bf X}^\lp\cdot {\bf P}_0 ) \,, \nn\\
&&\{ Q^+_{(0)\,\a }, \bar{Q}^+_{(0)\,\b }\, \} =  \label{w3} \\
&&\hspace{12mm} (\g_+)_{\a\b}\,\Big(\, {\bf P}_0\cdot{\bf P}_{0} 
+\ft{1}{2}\, 
(f_{\l\lp 0}\,X^{a\, \l}\, X^{b\, \lp})^2\Big) \nn\\
&& \hspace{12mm}
 +2\, (\g_a\,\g_+ )_{\a\b}\, f_{\l\lp 0}\, X^{a\,\l}\, {\bf X}^\lp\cdot 
{\bf P}_0  \,, \nn\\
&& \{ Q^+_{(0)\,\a } , \bar{Q}^-_\b\, \} = -(\g_a\,\g_+\,\g_-)_{\a\b}\,
P^a_0  \label{s4}\\
&& \hspace{12mm}- \ft{1}{2}\, (\g_{ab}\,\g_+\, \g_-)_{\a\b}\, 
f_{\l\lp 0}\, X^{a\, \l}\, X^{b\, \lp }\, , \nn\\
&& \{ \whQ^+_\a, \bar{Q}^-_\b\, \} = \{\, Q^+_{(0)\,\a } \, , 
\bar{\widehat{Q}}{}^+_\b\, \} = 0 \,. 
\la{modesusy}
\eea
And the mass operator commutes with all the supersymmetry charges,
\be
[\, \whQ^+, {\cal M}^2\, ] =[\, Q_{(0)}^+, {\cal M}^2\, ] =
[\, Q^-, {\cal M}^2\, ] = 0 ,
\ee{QswithM2}
defining a supersymmetric quantum-mechanical model. 

At this stage it would be desirable to present a matrix
model regularization of the supermembrane with winding contributions,
generalizing the matrix approximation to the exact subgroup of
area-preserving diffeomorphisms \cite{GoldstoneHoppe,dWHN}, at
least for toroidal 
geometries. However, this program seems to fail due to the fact that the 
finite-$N$ approximation to the structure constants \tmath{f_{\l BC}} 
violates the Jacobi identity, as was already noticed in \cite{dWMN}. 

Let us nevertheless discuss an example of this regularization in the
case of non-winding, open supermembranes in some detail. Consider 
a spacesheet topology of an annulus. Its set of basis functions
is easily obtained by starting from the well-known torus functions 
$Y_{\bf m}(\sigma)$ \cite{dWHN}
labeled by a two-dimensional vector ${\bf m}=(m_1,m_2)$ with $m_1,m_2$
integer numbers and
\be
Y_{\bf m}(\sigma_1,\sigma_2)=e^{i\, (m_1\sigma_1+m_2\sigma_2)}.
\ee{Ys}
Consider now the involution ${\bf m}\rightarrow {\bf \widetilde{m}}$
where ${\bf \widetilde{m}}=(m_1, -m_2)$.
One then defines the new basis functions \cite{kimrey,EMM2}
\be
C^\pm_{\bf m}= Y_{\bf m} \pm Y_{\bf \widetilde{m}},
\ee{Ces}
where $\s_1\in[0,2\pi]$ and $\s_2\in[0,\pi]$.
It turns out that $C^+_{\bf m}$ obeys
Neumann  and $C^-_{\bf m}$ Dirichlet conditions on the 
boundaries, i.e., $\del_2C^+_{\bf m}|=\del_1C^-_{\bf m}|=0$. These
basis functions possess the algebra
\bea
\{ C^-_{\bf m},C^-_{\bf n} \}&=& i ({\bf m}\! \times\! {\bf n})\, C^-_{\bf m+n}
-i({\bf m} \!\times\! {\bf \widetilde{n}})\, C^-_{\bf m+\wtd{n}}\,, \nn\\
\{ C^+_{\bf m},C^+_{\bf n} \}&=& i ({\bf m}\! \times\! {\bf n})\, C^-_{\bf m+n}
+i({\bf m} \!\times\! {\bf \widetilde{n}})\, C^-_{\bf m+\wtd{n}}\,,\nn \\
\{ C^+_{\bf m},C^-_{\bf n} \}&=& i ({\bf m}\! \times\! {\bf n})\, C^+_{\bf m+n}
-i({\bf m} \!\times\! {\bf \widetilde{n}})\, C^+_{\bf m+\wtd{n}} \, .
\nn\\&&\la{Calg}
\eea
Note that the $C^-_{\bf m}$ form a closed subalgebra.

The matrix regularization now comes about by replacing $Y_{\bf m}$
with the $(N^2-1)$ adjoint SU($N$) matrices $T_{\bf m}$ of \cite{dWHN}. 
The corresponding
operation to ${\bf m}\rightarrow\widehat{\bf m}$ is matrix transposition, i.e.
$T_{\bf \wtd{m}}=T_{\bf m}^{\rm T}$. Hence we find that, in the
matrix picture, the antisymmetric $(N(N-1)/2)$ $C^-_{\bf m}$ matrices
form the adjoint and  the symmetric
$(N(N+1)/2)$ $C^+_{\bf m}$ transform as the symmetric rank-two representation
of SO($N$).

Finally we turn to the question of the mass spectrum for 
membrane states with winding. The mass spectrum 
of the supermembrane without winding is continuous. This was 
proven in the SU($N$) regularization \cite{dWLN}. Whether or 
not nontrivial zero-mass states exist, is not known (for some 
discussion on these questions, we refer the reader to 
\cite{DWN}). Those would coincide with the states of
11-dimensional supergravity. It is often   
argued that the winding may remove the continuity of the 
spectrum (see, for instance, \cite{Russo}). 
There is no question that winding may increase the 
energy of the membrane states. A membrane winding around more 
than one compact  
dimension gives rise to a nonzero central charge in the 
supersymmetry algebra. This central charge sets a lower limit on  
the membrane mass. 
However, this should not be interpreted as an indication that 
the spectrum becomes discrete. The possible continuity of the spectrum 
hinges on two features. First the system should possess
continuous valleys of classically degenerate states. 
Qualitatively one recognizes immediately that this feature is not 
directly affected by the winding. A classical membrane with 
winding can still have stringlike configurations of arbitrary length,  
without increasing its area. Hence the classical instability 
still persists. 

The second feature is supersymmetry. Generically the classical 
valley structure is lifted by quantum-mechanical corrections, 
so that the wave function cannot escape to infinity. This 
phenomenon can be 
understood on the basis of the uncertainty principle. Because, at 
large distances, the valleys become increasingly narrow, the wave 
function will be squeezed more and more which tends to induce an 
increasing spread in its momentum. This results in an increase of 
the kinetic energy. Another way to understand this is by noting 
that the transverse oscillations perpendicular to the valleys 
give rise to a zero-point energy, which acts as an effective 
potential barrier that confines the wave function. When the 
valley configurations are supersymmetric the contributions from 
the bosonic and the fermionic transverse oscillations cancel each 
other, so that the wave function will not be confined and can
extend arbitrarily far into the valley. This phenomenon indicates
that the energy spectrum must be continuous. 

Without winding it is clear that the valley configurations are 
supersymmetric, so that one concludes that the spectrum is 
continuous. With winding the latter aspect is somewhat more 
subtle. However, we note that, when the winding density is
concentrated in one part of the spacesheet, then valleys can
emerge elsewhere
corresponding to stringlike configurations with supersymmetry.
Hence, as a space-sheet local field theory,  
supersymmetry can be broken in one region where the winding is 
concentrated and unbroken in 
another. In the latter  region stringlike configurations can form, 
which, at least semiclassically, will not be suppressed by 
quantum corrections. 

We must stress that we are describing only the 
generic features of the spectrum. Our arguments by no means 
preclude the existence of mass gaps. To prove or disprove the 
existence of  
discrete states is extremely difficult. While the contribution of 
the bosonic part of the Hamiltonian increases by concentrating the 
winding density on part of the spacesheet, the matrix elements in 
the fermionic directions will also grow large, making it 
difficult to estimate the eigenvalues. At this moment the only 
rigorous result is the BPS bound that follows from the supersymmetry 
algebra. Obviously, the state of  lowest mass for 
given winding numbers, is always a BPS state, which is invariant under some
residual supersymmetry.

%%%%%%%%%%%%%%%%%%%%%%%%%%%%%%%%%%%%%%%%%%%%%%%%%%%%%%%%%%

\end{document}